\documentclass[draft,final]{elsart}

\usepackage{amsmath,amssymb}
\usepackage[dvips]{graphicx}

\renewcommand{\eqref}[1]{eq.~(\ref{#1})}
\newcommand{\eqsref}[1]{eqs.~(\ref{#1})}
\newcommand{\pref}[1]{(\ref{#1})}
\renewcommand{\(}{\left(}
\renewcommand{\)}{\right)}

\newcommand{\p}{\partial}
\newcommand{\expct}[1]{\langle #1 \rangle}
\newcommand{\err}[3]{#1^{+(#2)}_{-(#3)}}
\newcommand{\Err}[3]{$#1^{+(#2)}_{-(#3)}$}
\DeclareMathOperator{\sgn}{sgn}

\begin{document}

\begin{frontmatter}



\title{Can the Ising critical behavior survive in non-equilibrium synchronous cellular automata?}


\author{Kazumasa Takeuchi\corauthref{Takeuchi}}
\address{Department of Physics, The University of Tokyo, 7-3-1 Hongo, Bunkyo-ku, Tokyo, 113-0033, Japan}
\corauth[Takeuchi]{Corresponding author. \\ Tel. and fax : +81-3-5841-4183, \\ E-mail address : kazumasa@daisy.phys.s.u-tokyo.ac.jp}
% \ead{kazumasa@daisy.phys.s.u-tokyo.ac.jp}

\begin{abstract}
Universality classes of Ising-like phase transitions are investigated
 in series of two-dimensional synchronously updated
 probabilistic cellular automata (PCAs),
 whose time evolution rules are either of Glauber type
 or of majority-vote type, and degrees of anisotropy are varied.
Although early works showed that coupled map lattices and PCAs
 with synchronously updating rules belong to a universality class
 distinct from the Ising class,
 careful calculations reveal that
 synchronous Glauber PCAs should be categorized
 into the Ising class, regardless of the degree of anisotropy.
Majority-vote PCAs for the system size considered
 yield exponents $\nu$ which are between those of the two classes,
 closer to the Ising value, with slight dependence on the anisotropy.
The results indicate that the Ising critical behavior is robust
 with respect to anisotropy and synchronism
 for those types of non-equilibrium PCAs.
There are no longer any PCAs known to belong to the non-Ising class.
\end{abstract}

\end{frontmatter}

% main text
\section{Introduction}

%There is no doubt that the notion of phase transition
% and universality class has been one of the central issues
% in equilibrium physics.
%Roughly speaking, the divergence of correlation length and time
% at second-order transitions makes it possible to classify
% diverse phenomena in terms of a few universality classes.
%Since being at equilibrium itself does not seem to be essential there,
% the concept of universality is widely believed to hold
% in some range of non-equilibrium systems
% \cite{Odor,Hinrichsen}.
The concept of universality class, which has been undoubtedly
 one of the central issues in equilibrium physics,
 is widely believed to hold in some range of non-equilibrium systems
 \cite{Odor,Hinrichsen}.
That is, even far from equilibrium, many microscopic details of systems
 become irrelevant at transition points,
 and a set of critical exponents depends
 only on a small number of basic, macroscopic ingredients.
The ``basic ingredients'' comprise, e.g., spatial dimension,
 symmetries and conservation laws, as in equilibrium systems,
 but then it is natural to ask
 ``do there exist any additional relevant parameters
 intrinsic to non-equilibrium?''
% which affect universality classes?''

The Ising universality class
%, which is a prototype of equilibrium classes,
 is also observed in non-equilibrium systems.
For example, probabilistic cellular automata (PCAs) and
 coupled map lattices (CMLs) with up-down symmetry can exhibit
 Ising-like transitions by varying a control parameter
 such as a coupling constant.
Grinstein \textit{et al.} \cite{Grinstein_etal} showed that,
 suppose a coarse-grained dynamics of such systems is described
 by a Langevin equation,
% with a Gaussian white noise,
 irreversibility due to the broken detailed balance is irrelevant
 and the considered models should fall in the Ising class,
 using the standard dynamic renormalization group treatment.
This prediction has been actually confirmed
% both numerically and theoretically
 in various kinetic Ising models with asynchronously updating rules
 \cite{KineticIsing}.
% \cite{Odor}.
However, Marcq \textit{et al.} \cite{Marcq_etal-1} numerically found that
 Ising-like transitions of some two-dimensional non-equilibrium CMLs separate
 into \textit{two} distinct universality classes: 
 the Ising class, and a new universality class,
 called ``non-Ising class'' hereafter,
 where the correlation length exponent $\nu$ is $0.90 (2)$
 [here number(s) between parentheses indicate the uncertainty
 in the last digit(s) of the quantity],
 which differs from the Ising value $\nu = 1$.
Ratios of exponents $\beta/\nu$ and $\gamma/\nu$
 are common to the two classes, namely $0.125$ and $1.75$, respectively.
Since CMLs with synchronously updating rules form the new class,
 whereas asynchronously updated ones belong to the Ising class,
 synchronism is thought to be a relevant parameter.
After their work, several synchronous systems such as
 stochastic CMLs \cite{Sastre_etal},
 a logistic CML at the onset of a non-trivial collective behavior
 \cite{Marcq_etal-2},
 and even PCAs \cite{Makowiec,Makowiec_etal}
 have been investigated and
 the existence of the non-Ising class has been observed again,
 aside from a few exceptions \cite{Sastre_etal,Perez_etal}.
Thus the importance of synchronism,
% of updating schemes,
 which may be related to the existence of an external clock,
 has been attracted much attention
% and recently it extends to the effect on stabilization
 \cite{Stabilization}.

However, it is obvious that the synchronous updating
 does \textit{not} immediately bring about the non-Ising critical behavior:
 there exist synchronous PCAs which respect the detailed balance
 and therefore we can safely say that they belong to the Ising class,
 thanks to the equilibrium universality hypothesis.
For example, an isotropic PCA with a Glauber transition rate
 satisfies that condition.
On the other hand, a Glauber PCA with completely anisotropic interaction
 was numerically studied by Makowiec and Gnaci\'{n}ski (MG)
 \cite{Makowiec_etal}, and concluded to be in the non-Ising class.
The above two observations naturally lead to a supposition that
 a degree of anisotropy may affect the selection of the universality classes.
The aim of this paper is, therefore,
 to examine the relation between anisotropy and the universality classes.

\section{Models}

Two series of PCAs with up-down symmetry,
 namely Glauber PCAs and majority-vote PCAs,
 are numerically investigated.
% for the above purpose.
Both of them are on a two-dimensional,
 square lattice of size $L \times L$
 with periodic boundary conditions.
A local variable, or ``spin,'' $s_{i,j}^t \in \{-1,+1\}$
 is assigned to each lattice point,
 where indices $i$ and $j$ denote Cartesian coordinates,
 and $t$ is the discrete time.
%The definitions of PCAs are completed with time evolution rules.
Each site $(i,j)$ is simultaneously updated
 according to a specific local transition probability
 $p_{i,j}(\pm 1|\{s\})$ that the spin takes a value $\pm 1$
 after one time step from a spin configuration $\{s\}$.
For the Glauber PCAs, it is defined as
\begin{equation}
 p_{i,j}(\pm 1|\{s\}) = \frac{1}{2} \{ 1 \pm \tanh g [ s_{i,j}+s_{i+1,j}+s_{i,j-1} +\alpha ( s_{i-1,j}+s_{i,j+1} )] \},  \label{eq:2.1}
\end{equation}
 where $g$ is a coupling constant acting as a control paramter,
 and $0 \leq \alpha \leq 1$ indicates the degree of anisotropy.
We can realize a variety of Glauber PCAs with different degrees of anisotropy
 by varying the value of $\alpha$.
An isotropic case corresponds to $\alpha = 1$,
 in which it is easily shown that the detailed balance is satisfied
 with stationary distribution
%\begin{equation}
 $P(\{s\}) \equiv \frac{1}{Z} \e^{-\mathcal{H}(\{s\})},
 \mathcal{H}(\{s\}) = -\sum_{i,j} \ln \cosh g \( s_{i,j}+s_{i-1,j}+s_{i+1,j}+s_{i,j-1}+s_{i,j+1} \),$
%  \label{eq:2.2}
%\end{equation}
 where $\mathcal{H}(\{s\})$ is an effective Hamiltonian and
 $1/Z$ is a normalization constant.
Since the Hamiltonian respects the up-down symmetry
 and consists only of short-range interactions,
 the equilibrium universality hypothesis asserts that
 this isotropic Glauber PCA should fall in the Ising class
 despite the synchronous updating scheme.
On the other hand, the detailed balance does not hold for $\alpha \neq 1$.
In particular, the completely anisotropic case $\alpha = 0$
 is already reported by MG to have the exponent $\nu = 0.93(3)$
 and to belong to the non-Ising class \cite{Makowiec_etal}.

The other series of PCAs is that of majority-vote PCAs, defined as
\begin{equation}
 p_{i,j}(\pm 1|\{s\}) = \frac{1}{2} \{ 1 \pm g \sgn [ s_{i,j}+s_{i+1,j}+s_{i,j-1} +\alpha ( s_{i-1,j}+s_{i,j+1} )] \}.  \label{eq:2.3}
\end{equation}
For this model, the degree of anisotropy is classified into only 3 types:
 completely anisotropic $0 \leq \alpha < 0.5$,
 intermediate $\alpha = 0.5$, and isotropic $0.5 < \alpha \leq 1$.
The detailed balance is violated in all of them
 \cite{Grinstein_etal,Lebowitz_etal}.
The completely anisotropic case is well-known as the Toom PCA \cite{Toom_etal},
% and MG also estimated its exponent to be $\nu = 0.86 (4)$,
 whose exponent was also estimated by MG to be $\nu = 0.86 (4)$,
 i.e. the non-Ising value \cite{Makowiec, Makowiec_etal}.

\section{Methodology}

All of the simulations are implemented
 on lattices of size up to $L=96$
% ranging from $L=16$ to $96$,
 by the following procedure.
We start from random initial conditions
 and discard first $t_0 = 10^5$ time steps as transients.
It is sufficiently long to consider the systems
 in the time-asymptotic attractor,
 since the correlation time is
% approximately
 $\tau_\text{corr} \sim 5 \times 10^3$ in the worst case.
%Eye observation and time series data also support it.
Then time series of the averaged spin
 $m_L^t \equiv (1/L^2) \sum_{i,j} s_{i,j}^t$
 is used to calculate the magnetization $M_L$,
 higher order moments $M_L^{(n)}$,
 the susceptibility $\chi_L$ and the Binder's cumulant $U_L$ \cite{Binder},
 defined by
\begin{align}
 &M_L = \expct{|m_L^t|},~~~~
 M_L^{(n)} = \expct{|m_L^t|^n},~~~~ \notag \\
 &\chi_L = L^2 (M_L^{(2)}-M_L^2),~~~~
 U_L(g) = 1-\frac{M_L^{(4)}(g)}{3M_L^{(2)}(g)^2},  \label{eq:3.1}
\end{align}
 where $\expct{\cdots}$ denotes the expectation value
 obtained by integrating an observable during the duration $T$, such as
 $\expct{|m_L^t|} = (1/T) \sum_{t=t_0+1}^{t_0+T} |m_L^t|$.
Absolute values of $m_L^t$ are taken in \eqsref{eq:3.1} as usual,
 since finite-size effects allow sign reversals
 of the instantaneous magnetization even in the ordered phase.
The integration time $T$ is $8 \times 10^7$ for the isotropic Glauber PCA
 and $1.5 \times 10^8$ for the others,
 and thus much longer than
 both the correlation time $\tau_\text{corr} \lesssim 5 \times 10^3$
 and the sign reversal time $\tau_\text{rev} \lesssim 5 \times 10^5$.
Therefore, the quantities in \eqsref{eq:3.1} are capable of representing
 the corresponding ensemble averages with good accuracy.

\section{Measurement of critical exponents}

Critical exponents $\beta, \gamma, \nu$ are estimated
 for the Glauber PCAs with $\alpha=0$, $0.25$, $0.5$, $0.75$, $1$
 and the majority-vote PCAs with $\alpha=0$, $0.5$, $1$.
The way of measurement is almost similar to
 that of Marcq \textit{et al.} and later works
 \cite{Marcq_etal-1,Sastre_etal,Makowiec,Makowiec_etal,Perez_etal},
 that is, to exploit finite-size scaling laws in equilibrium
 which are empirically known to hold
 in the non-equilibrium Ising-like transitions.
The following shows the process of measurement for the Glauber PCA
 with $\alpha = 0.75$ as a typical case.

\begin{figure}
 \begin{minipage}{.49\hsize}
  \begin{center}
   \includegraphics[width=\hsize]{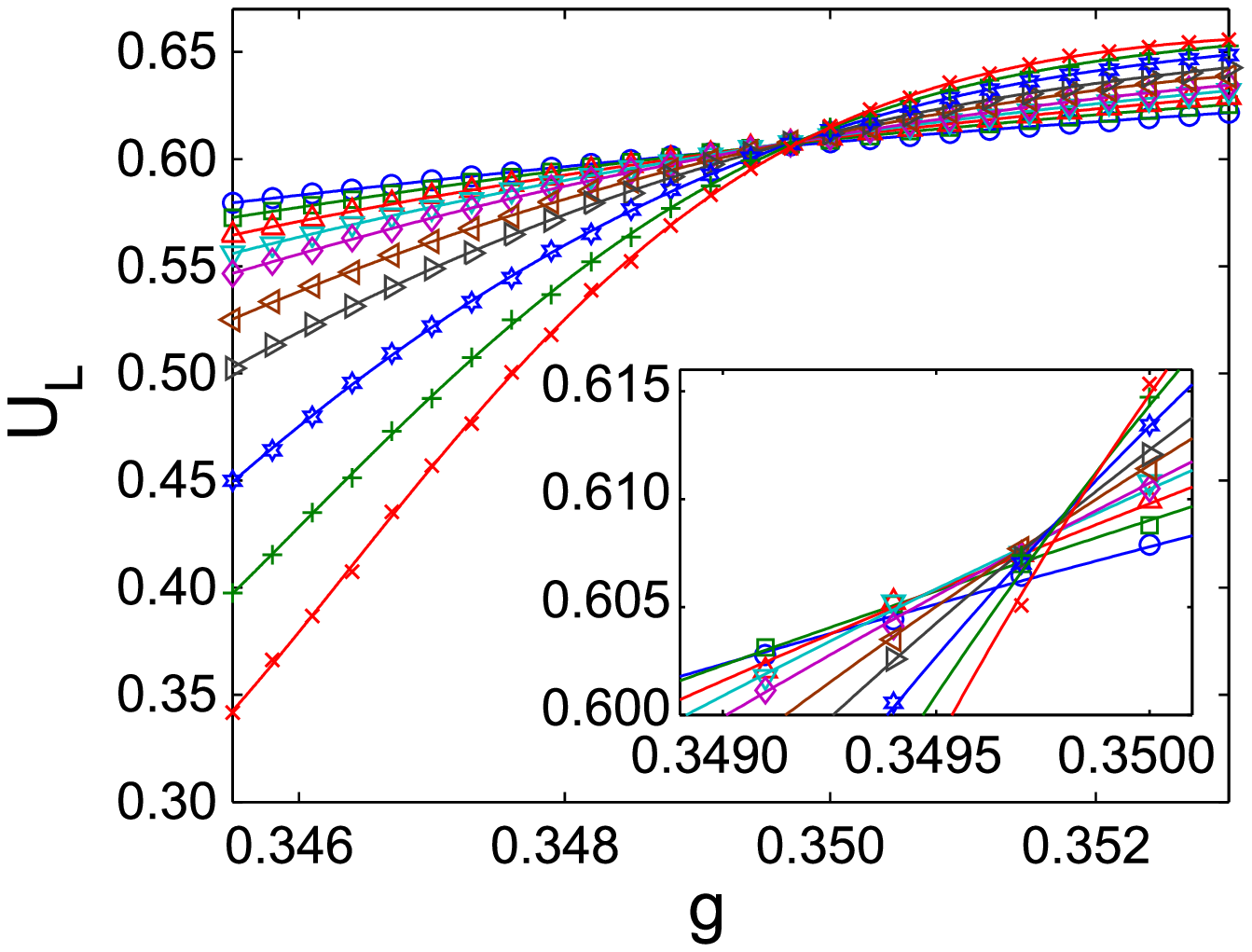}
   \caption{Plots of the Binder's cumulant $U_L(g)$ for the Glauber PCA with $\alpha = 0.75$. System sizes are, from the smallest slope, $L=16$, $20$, $24$, $28$, $32$, $40$, $48$, $64$, $80$, $96$. Symbols correspond to raw data and curves indicate 5th order polynomial fits. The inset is a magnification of the intersection region.}
   \label{fig:1}
  \end{center}
 \end{minipage}
 \hspace{.02\hsize}
 \begin{minipage}{.49\hsize}
  \begin{center}
   \includegraphics[width=\hsize,clip]{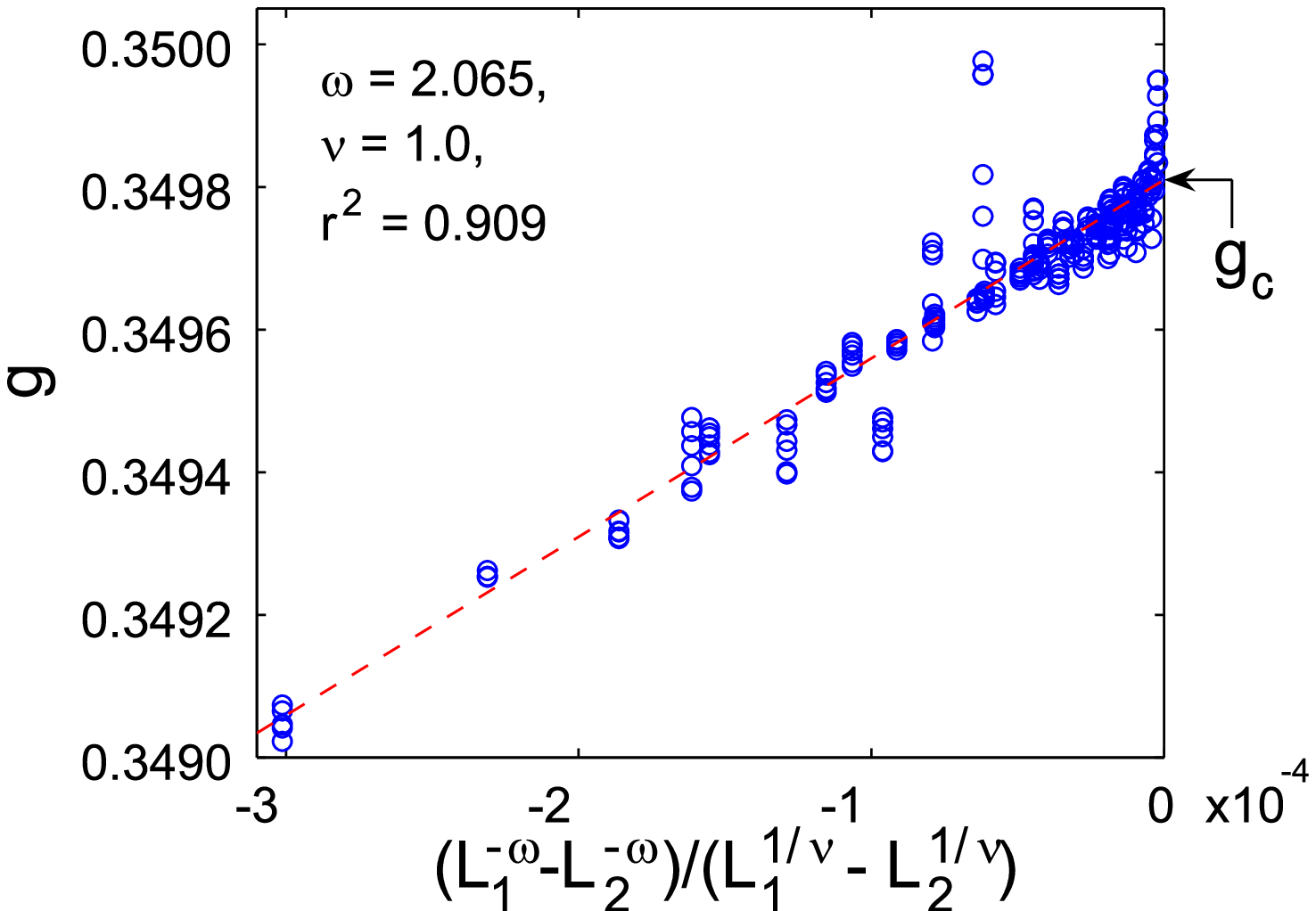}
   \caption{Plot of coordinates $g$ of intersections between curves $U_{L_1}(g)$ and $U_{L_2}(g)$. The intersections are located by fitting raw data with polynomials of orders $4$ to $9$, and superimposed on the same figure. The range of the orders is chosen so that the fitting curves adequately trace the raw data. The broken line indicates the linear regression, whose y-intercept yields an estimate of $g_c$. See \eqref{eq:4.2}.}
   \label{fig:2}
  \end{center}
 \end{minipage}
\end{figure}

The first to do is to find the critical point $g=g_c$.
In order to locate it, we adopt the standard method
 using the Binder's cumulant $U_L(g)$ \cite{Binder}.
%\begin{equation}
% U_L(g) = 1-\frac{M_L^{(4)}(g)}{3M_L^{(2)}(g)^2}.  \label{eq:4.1}
%\end{equation}
Since the cumulant has the scaling form of
 $U_L(g) = \hat{U}((g-g_c)L^{1/\nu})$,
 it becomes independent of $L$ at criticality,
 i.e. $U_L(g_c)=U^*$ for all $L$.
The quantity $U^*$ is also a universal number.
Figure \ref{fig:1} shows plots of $U_L(g)$ and their polynomial fitting curves
 for various system sizes.

Focussing on intersections between them,
 a drift toward larger values of $g$ is clearly observed
 (the inset of fig. \ref{fig:1}).
This is caused by remaining contributions of irrelevant operators,
 which exist in finite-size systems \cite{Binder}.
%Knowing the source of the errors enables us
% to improve the accuracy of estimation.
Taking it into consideration,
 the scaling function of the cumulant should be modified to
 $U_L(g) = \hat{U}((g-g_c)L^{1/\nu}, \mathcal{O} L^{-\omega})$,
 where only one irrelevant term
 with a relaxation exponent $\omega$ is assumed to be dominant.
An expansion of the modified scaling form in powers of $(g-g_c)L^{1/\nu}$
 and $L^{-\omega}$ up to the 1st order
 yields the coordinate of the intersection
 between $U_{L_1}(g)$ and $U_{L_2}(g)$, namely,
\begin{equation}
 g = g_c + A \frac{L_1^{-\omega} -L_2^{-\omega}}{L_1^{1/\nu}-L_2^{1/\nu}},~~~~
 U = U^* + B \frac{L_1^{1/\nu}L_2^{-\omega}-L_2^{1/\nu}L_1^{-\omega}}{L_1^{1/\nu}-L_2^{1/\nu}},  \label{eq:4.2}
\end{equation}
 where $A$ and $B$ are expansion coefficients.
Consequently, by plotting $g$ with respect to
 $(L_1^{-\omega} -L_2^{-\omega})/(L_1^{1/\nu}-L_2^{1/\nu})$
 for fixed, reasonable values of $\omega$ and $\nu$,
 and by searching for $\omega$
 which gives the highest correlation coefficient for it,
 we obtain estimates of both $\omega$ and $g_c$.
Figure \ref{fig:2} is a plot for $\nu=1$ and the best value of $\omega$,
 where we can clearly see the linear dependence,
% between the two quantities,
 which gives the coefficient of determination $r^2 = 0.909$.
The critical point $g_c$ is then estimated
 as a y-intercept of the regression line,
 $g_c = \err{0.34981}{4}{3}$ in this case.
Here, the superscript (subscript) indicates the confidence interval
 in the plus (minus) direction, which is evaluated as the region
 with sufficiently large coefficient of determination, namely,
 $(1-r^2) < (1-\max_\omega r^2) \times 1.1$.
The estimation of $U^*$ is carried out quite similarly,
 which results in $U^* = \err{0.6107}{41}{13}$.
This is in good agreement with the 2D-Ising value $U^* = 0.611 (1)$
 \cite{Privman_etal}.
A problem of this method may be
 that it requires the value of $\nu$, which is measured later, 
 but the difference in the estimates
 with respect to the assumed value of $\nu$,
 $1.0$ or $0.9$, turns out to be quite subtle:
 $2 \times 10^{-6}$ for $g_c$ and $2 \times 10^{-5}$ for $U^*$,
 i.e. negligible.
The mentioned improvement is therefore useful and 
 expected to lead to more reliable estimation of critical exponents.
% recalling that the distributed intersections of the cumulant had been
% one of the main sources of errors in it.

Now we proceed to the measurement of critical exponents.
The following finite-size scaling relations are made use of to achieve it:
\begin{subequations}
\begin{align}
 &\p_g \ln M_L|_{g_c} \sim L^{1/\nu}, &&\p_g \ln M_L^{(2)}|_{g_c} \sim L^{1/\nu}, \notag \\
 &\p_g \ln M_L^{(4)}|_{g_c} \sim L^{1/\nu}, &&\p_g U_L|_{g_c} \sim L^{1/\nu},  \notag \\
 &V_2 \equiv [m^2]^2/[m^4] \sim L^{1/\nu}, &&V_4 \equiv ([m]^4/[m^4])^{1/3} \sim L^{1/\nu},  \notag \\
 &V_6 \equiv [m]^2/[m^2] \sim L^{1/\nu},  \label{eq:4.3a}
\end{align}
 and
\begin{align}
 &M_L(g_c) \sim L^{-\beta/\nu}, &&M_L^{(2)}(g_c)^{1/2} \sim L^{-\beta/\nu}, \notag \\
 &M_L^{(4)}(g_c)^{1/4} \sim L^{-\beta/\nu}, &&\chi_L(g_c) \sim L^{\gamma/\nu},  \label{eq:4.3b}
\end{align}  \label{eq:4.3}
\end{subequations}
 where $[m^n] \equiv \p_g M_L^{(n)}|_{g_c}$,
 and the derivatives are evaluated
 by using polynomial fittings of appropriate orders.
Irrelevant operators can affect the scaling relations \pref{eq:4.3} again.
We, therefore, perform the measurement as follows.
(I) First, we plot a quantity in \eqsref{eq:4.3} in the log-log scale,
 using a polynomial fit of a fixed order, and check the linear dependence.
(IIa) If the contributions of irrelevant fields are already suppressed
 for the smallest size considered, $L=16$,
 a simple linear fit giving the lowest chi-square in the log-log plot
 yields an estimate.
(IIb) Otherwise, finite-size corrections are employed to achieve
 an asymptotic value of the slope,
 after Marcq \textit{et al.} and subsequent works
 \cite{Marcq_etal-1,Sastre_etal}.
An expansion is made similar to that used for $U_L(g)$ above,
 e.g. $L^{-1/\nu} \p_g \ln M_L|_{g_c} \simeq C_1 + C_2 L^{-\omega}$,
 followed by searching for the best $\omega$ and $\nu$ (or other exponents),
 in the sense of lowest chi-square.
This method, however, does not work sometimes due to rapid convergence
 of correction terms, or statistical errors in the raw data.
In that case, (IIc) we use a linear fit in the log-log plot
 neglecting data points subject to the finite-size effect.
Care is taken to ensure that the observed scaling behavior does
 indeed correspond to the asymptotic regime, in all of the cases.
Finally, (III) we repeat the aforesaid procedure
 for polynomial fits of different orders in (I),
 and also for all of the quantities in \eqsref{eq:4.3}
 which give the same exponent.

\begin{figure}
 \begin{center}
  \includegraphics[width=.48\hsize,clip]{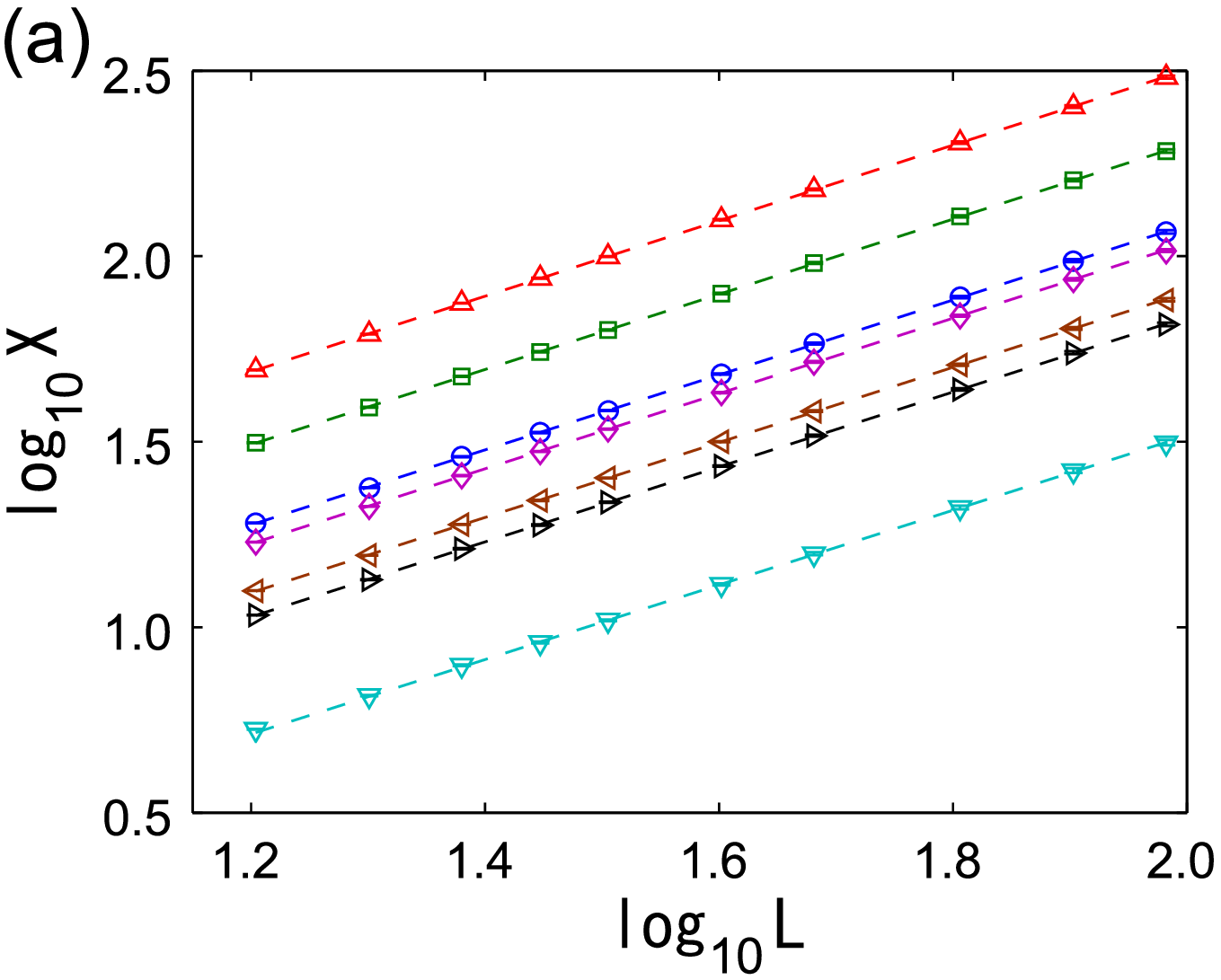}
  \hspace{.01\hsize}
  \includegraphics[width=.48\hsize,clip]{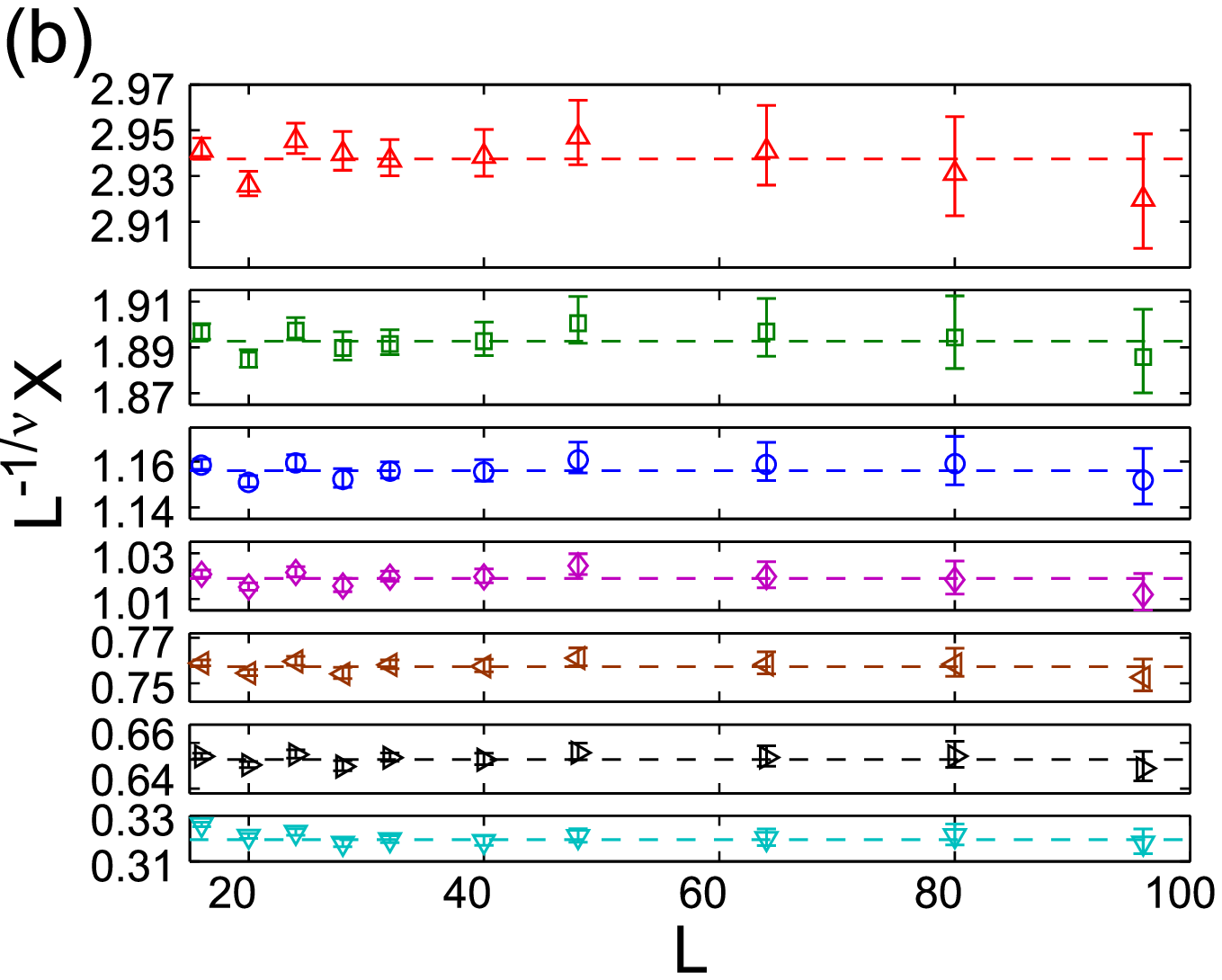}
 \end{center}
 \caption{(a) Scaling relations of the 7 quantities in \eqref{eq:4.3a} for the Glauber PCA with $\alpha=0.75$, where $X=\p_g \ln M_L^{(4)}|_{g_c}$, $\p_g \ln M_L^{(2)}|_{g_c}$, $\p_g \ln M_L|_{g_c}$, $V_2$, $V_4$, $V_6$, $\p_g U_L|_{g_c}$ from top to bottom. Corresponding slopes are $1.017$, $1.013$, $1.011$, $1.014$, $1.012$, $1.011$ and $1.006$, respectively. (b) Deviations from the scaling laws, which are found to be faint. Note that the scale in the y-axis is identical for all of the subplots.}
 \label{fig:3}
\end{figure}

A value of $\nu$ for the Glauber PCA with $\alpha = 0.75$ is thereby
 evaluated to be $\nu = \err{0.988}{19}{15}$, as is shown in fig. \ref{fig:3}.
The main sources of errors are
 the uncertainty in the estimate of the critical point $g_c$ and
 statistical errors due to the finite-time sampling.
The former is estimated from the confidence interval of $g_c$,
 while the latter is from the dependence of exponents
 on the quantity used for the finite-size scaling,
 and on the degree of the fitting polynomials.
Note that the final range of errors
% in the estimates of exponents
 is given as the union of each error region considered,
 i.e. as wide as possible.
Recalling the value of $\nu$ for the Ising and non-Ising class,
 $1$ and $0.90(2)$ respectively,
 we can clearly conclude that the Glauber PCA with $\alpha = 0.75$
 belongs to the Ising class.
Values of $\beta/\nu$ and $\gamma/\nu$ are obtained in the same way,
 which are $\beta/\nu = \err{0.1271}{17}{28}$ and
 $\gamma/\nu = \err{1.749}{15}{14}$,
 and again completely consistent with the Ising values.

\section{Results and discussions}

Critical exponents of all of the models considered
 are measured in the same manner,
 except for the following two points.
%(i) The number of $g$ per each size $L$ ranges from $22$ to $35$.
(i) Systems of size
 $L=16$, $20$, $24$, $28$, $32$, $40$, $48$, $64$, $80$, $96$ are examined
 for the Glauber PCAs with $\alpha=0.25$ and $0.75$,
 while those of size
 $L=16$, $24$, $32$, $40$, $48$, $64$, $80$, $96$
 are considered for the others,
 since finite-size corrections are appropriately caught by them.
(ii) Since no finite-size effects are observed in intersections of $U_L(g)$
 for the Glauber PCA with $\alpha=0$, values and errors of $g_c$ and $U^*$ are
 determined by means and standard deviations
 of the coordinates of the intersections.

\begin{figure}[b]
 \begin{minipage}{.49\hsize}
  \begin{center}
   \includegraphics[width=\hsize]{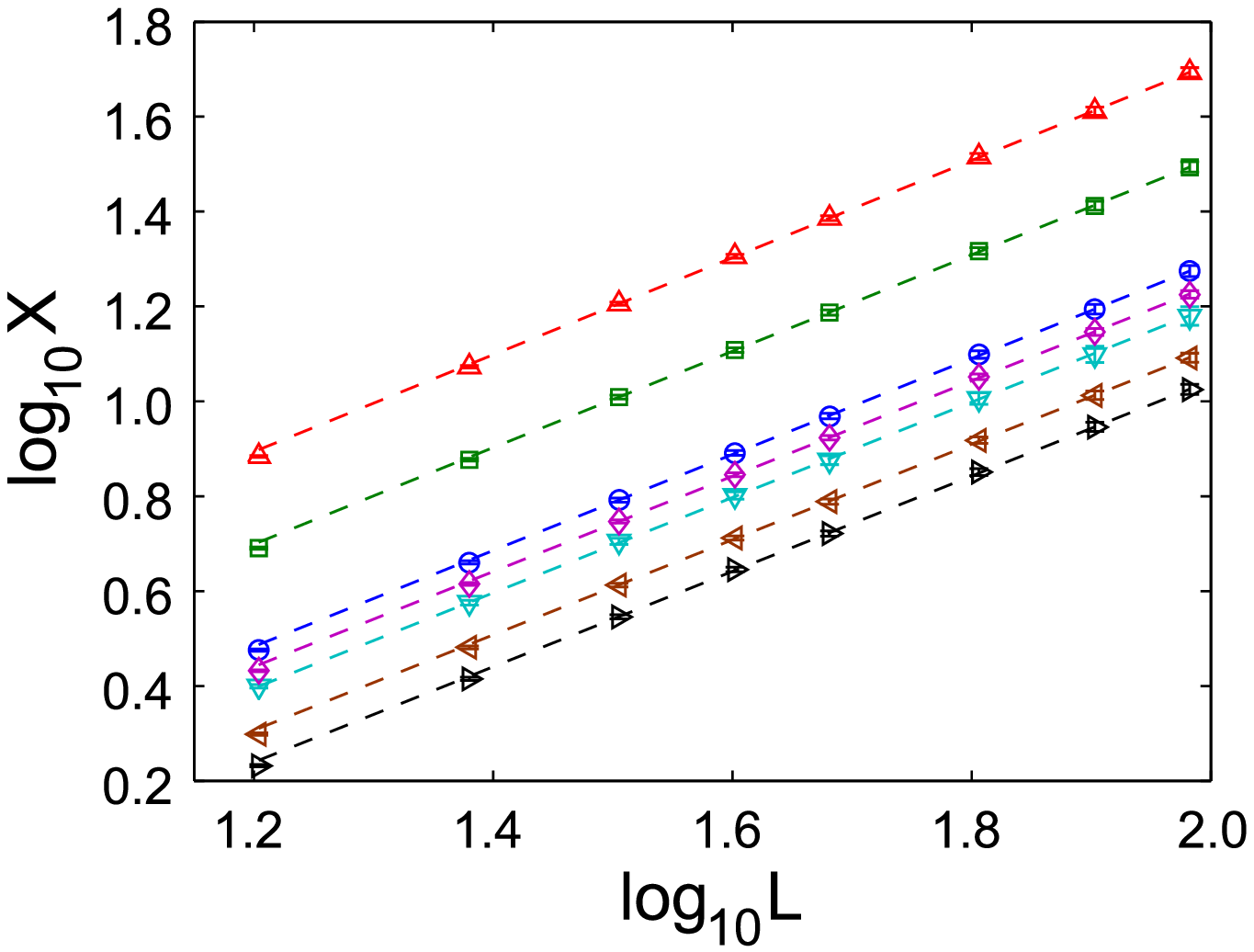}
   \caption{Measurement of the critical exponent $\nu$ for the Glauber PCA with $\alpha=0$. Each symbol and line corresponds to $X=\p_g \ln M_L^{(4)}|_{g_c}$, $\p_g \ln M_L^{(2)}|_{g_c}$, $\p_g \ln M_L|_{g_c}$, $V_2$, $\p_g U_L|_{g_c}$, $V_4$, $V_6$ from top to bottom. Slopes are $1.023$, $1.016$, $1.012$, $1.005$, $1.005$, $1.005$ and $1.005$, respectively.}
   \label{fig:4}
  \end{center}
 \end{minipage}
 \hspace{.02\hsize}
 \begin{minipage}{.49\hsize}
  \begin{center}
   \includegraphics[width=\hsize,clip]{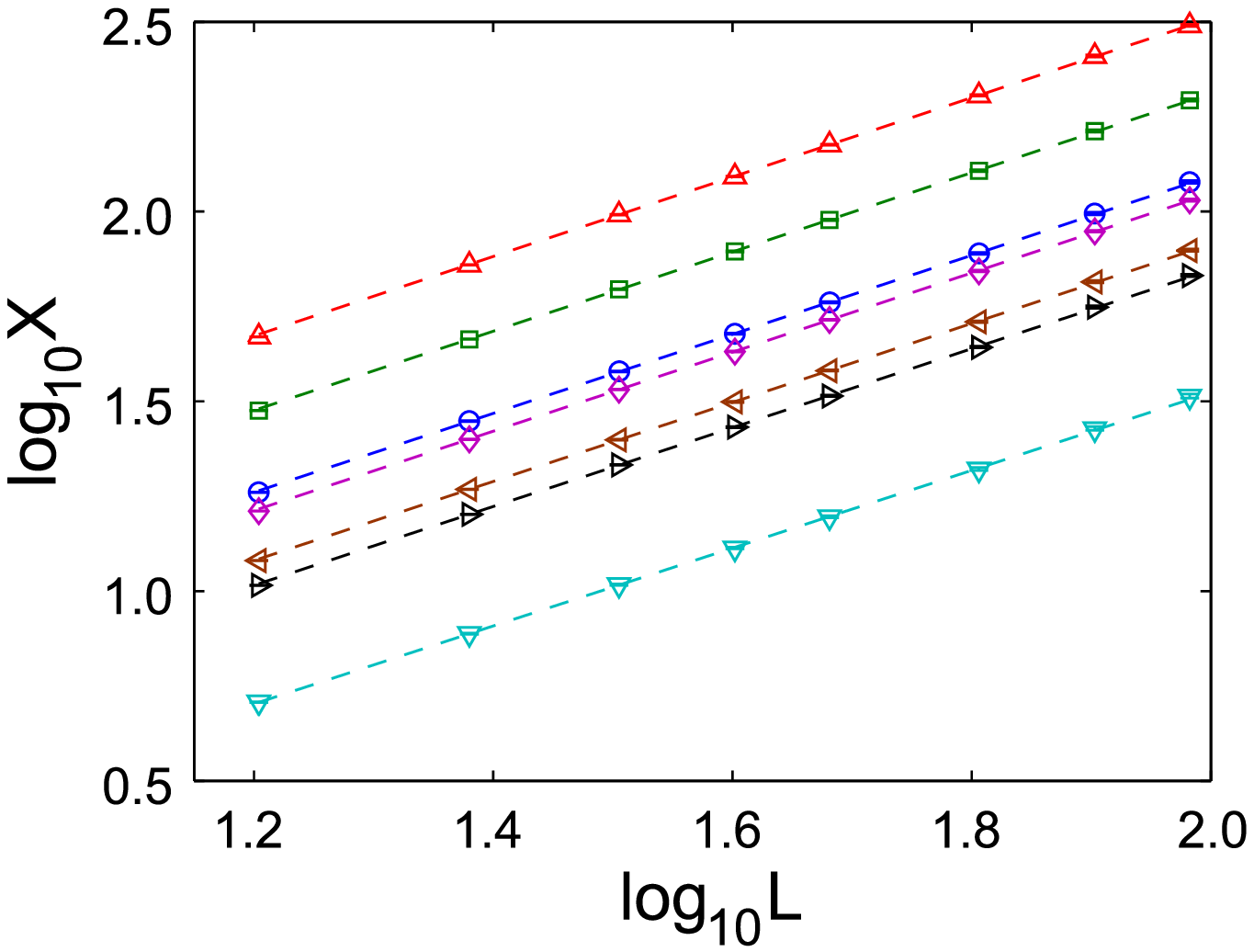}
   \caption{Measurement of the critical exponent $\nu$ for the majority-vote PCA with $\alpha=0$. Each symbol and line corresponds to $X = \p_g \ln M_L^{(4)}|_{g_c}$, $\p_g \ln M_L^{(2)}|_{g_c}$, $\p_g \ln M_L|_{g_c}$, $V_2$, $V_4$, $V_6$, $\p_g U_L|_{g_c}$ from top to bottom. Slopes are $1.046$, $1.044$, $1.041$, $1.044$, $1.041$, $1.039$ and $1.026$, respectively.}
   \label{fig:5}
  \end{center}
 \end{minipage}
\end{figure}

The measured exponents are summarized in table \ref{tbl:1}.
%All of the values of $U^*, \beta/\nu$ and $\gamma/\nu$ are consistent with
% those of the Ising model, similarly to other studies on the non-Ising class
% \cite{Marcq_etal-1,Sastre_etal,Marcq_etal-2}.
Our results for the Glauber PCAs clearly show that
 they fall into the Ising class.
While the isotropic case $\alpha=1$, which is at equilibrium,
 should actually be there,
 the results for all the anisotropic Glauber PCAs are quite unexpected
 since they are non-equilibrium, synchronous PCAs.
In particular, we obtain a value of the exponent
 for the completely anisotropic case $\alpha=0$
 as $\nu = \err{0.992}{28}{26}$ (fig. \ref{fig:4}),
 which conflicts with the estimate by MG,
 $\nu = 0.93(3)$ \cite{Makowiec_etal}.

\begin{table}
 \caption{Critical points and exponents of the PCAs examined.}
 \label{tbl:1}
 \begin{center}
  \begin{tabular}{|ll|lllll|} \hline
   \multicolumn{1}{|c}{model} & \multicolumn{1}{c|}{$\alpha$} & \multicolumn{1}{c}{$g_c$} & \multicolumn{1}{c}{$U^*$} & \multicolumn{1}{c}{$\beta/\nu$} & \multicolumn{1}{c}{$\gamma/\nu$} & \multicolumn{1}{c|}{$\nu$} \\ \hline
   & $1$ &\Err{0.31173}{6}{3}&\Err{0.6098}{14}{5}&\Err{0.1271}{17}{54}&\Err{1.757}{11}{28}&\Err{0.993}{48}{18} \\
   & $0.75$ &\Err{0.34981}{4}{3}&\Err{0.6107}{41}{13}&\Err{0.1271}{17}{28}&\Err{1.749}{15}{14}&\Err{0.988}{19}{15} \\
   Glauber & $0.5$ &\Err{0.40407}{6}{3}&\Err{0.6100}{20}{7}&\Err{0.1254}{18}{42}&\Err{1.757}{7}{26}&\Err{0.994}{24}{15} \\
   & $0.25$ &\Err{0.49049}{5}{3}&\Err{0.6097}{6}{2}&\Err{0.1266}{10}{21}&\Err{1.753}{10}{11}&\Err{0.994}{22}{16} \\
   & $0$ &\Err{0.65855}{16}{16}&\Err{0.6123}{17}{16}&\Err{0.1239}{32}{32}&\Err{1.753}{11}{12}&\Err{0.992}{28}{26} \\ \hline
   & $1$ &\Err{0.73151}{4}{4}&\Err{0.6105}{7}{4}&\Err{0.1247}{15}{17}&\Err{1.753}{12}{9}&\Err{0.979}{27}{17} \\
   majority-vote & $0.5$ &\Err{0.783324}{21}{17}&\Err{0.6104}{4}{3}&\Err{0.1260}{25}{12}&\Err{1.753}{7}{7}&\Err{0.962}{24}{10} \\
   & $0$ &\Err{0.82248}{3}{2}&\Err{0.6140}{29}{14}&\Err{0.1249}{18}{24}&\Err{1.740}{10}{11}&\Err{0.956}{28}{15} \\ \hline
   \multicolumn{2}{|l|}{Ising \cite{Privman_etal}}&&$0.611(1)$&$0.125$&$1.75$&$1$ \\
   \multicolumn{2}{|l|}{non-Ising \cite{Marcq_etal-1,Sastre_etal}}&&$\approx 0.611$&$\approx 0.125$&$\approx 1.75$&$0.90(2)$ \\ \hline
  \end{tabular}
 \end{center}
\end{table}

On the other hand, the results for the majority-vote PCAs are rather unclear:
 the estimated exponents $\nu$ are between the Ising and non-Ising value.
They are closer to the Ising exponent
 and this tendency slightly increases with isotropy.
There seem to be two possible interpretations for it.
One is that the critical exponents of the majority-vote PCAs
 depend continuously on system parameters,
 anisotropy in our demonstrations.
This is, however, commonly understood as an exceptional case,
 at least in equilibrium.
Instead, it is rather natural to consider that
 what we see here is a part of an extremely slow convergence
 in finite-size scalings toward the Ising asymptotic behavior,
 which is estimated to be reached for $L \sim O(10^4)$.
The slight dependence on the degree of anisotropy is then
 related to the difference in relaxation constants,
 which may be attributed to coupling intensity between sites,
 and/or an additional length scale caused by some hidden coherent structures
 such as in \cite{Grassberger_Schreiber-1}, if they exist.
In any case, further studies are essential to give a conclusion on it.
% though a system size required for convergence to the asymptotic behavior
% is estimated as $L \sim O(10^4)$ for the majority-vote PCA with $\alpha=0$,
% which is far from the reach of current computing power.
In addition, the estimate for the completely anisotropic majority-vote PCA,
 or the Toom PCA,
 $\nu = \err{0.956}{28}{15}$ (fig. \ref{fig:5}) is again incompatible with
 the value from MG, $\nu = 0.86(4)$ \cite{Makowiec,Makowiec_etal}.

The discrepancies between our estimates and MG's
 are caused by a few differences in steps toward estimation.
One is in the sampling method.
We started from random initial conditions
 and sampled $T=1.5 \times 10^8$ consecutive data
 after discarding first $t_0 = 10^5$ time steps,
 whereas MG chose ordered initial conditions, in which all spins are $+1$,
 and set $t_0 \leq 10^4, T = 10^4$
 as the price for repeating independent simulations not more than $5500$ times.
Since $t_0$ and $T$ of MG are in many cases
 comparable with the correlation time
 $\tau_\text{corr} \lesssim 5 \times 10^3$,
 and comparable with or much shorter than the sign reversal time
 $\tau_\text{rev} \lesssim 5 \times 10^5$,
 we consider that the sampling in MG is statistically insufficient
 and the influence from the ordered initial conditions may remain.
Another origin of the discrepancies
 is a way to determine transition points $g_c$.
We estimated them solely from the crossings of the cumulant $U_L(g)$,
 with finite-size corrections if possible,
 while MG first made a guess from the crossings and then
 determined them by searching
 for $g_c$ which gives the ratios of exponents $\beta/\nu$ and $\gamma/\nu$
 identical to those of the Ising class.
This is, however, rather perilous since estimates of critical exponents
 are very sensitive to various errors, including statistical ones.
Moreover, in general, a small uncertainty in critical points $g_c$
 leads to much larger errors in critical exponents.
In fact, if we assume the value of $g_c$ in MG
 for the Glauber PCA with $\alpha=0$,
 namely $g_c=0.6580$, we reproduce their result $\nu \approx 0.93$.
For the reasons above, we believe our results are more reliable.

In conclusion, we have investigated the Ising-like phase transitions
 in synchronous PCAs, namely the Glauber PCAs and the majority-vote PCAs,
 with different degrees of anisotropy.
Our calculations reveal that the Glauber PCAs belong to the Ising class
 regardless of the degree of anisotropy,
% while the majority-vote PCAs for the system size considered
% yield exponents $\nu$
% which are between those of the Ising and non-Ising classes,
% closer to the Ising value, with slight dependence on the anisotropy.
 and the majority-vote PCAs are also expected to do so,
 though the latter remains to be clearly shown.
The results indicate that the Ising critical behavior is robust
 with respect to anisotropy and synchronism for the PCAs,
 which coincide with the theoretical prediction of Grinstein \textit{et al.}
 \cite{Grinstein_etal} and observations by Sastre and P\'{e}rez
 where some degree of deterministic dynamics is required to bring about
 the non-Ising critical behavior \cite{Sastre_etal}.

\begin{ack}
The author, K. Takeuchi, gratefully acknowledges fruitful discussions with
 M. Sano, A. Awazu, H. Chat\'{e}, P. Marcq and H. Tasaki.
The author would also like to thank K. Nakajima
 for letting him use the PC cluster Cenju for this work.
\end{ack}

\end{document}